\documentclass[12pt,preprint]{aastex}

\newcommand{\chandra}{{\sl Chandra\/}}
\newcommand{\xte}{{\sl RXTE\/}}
\newcommand{\asca}{{\sl ASCA\/}}
\newcommand{\rosat}{{\sl ROSAT\/}}
\newcommand{\exosat}{{\sl EXOSAT\/}}
\newcommand{\einstein}{{\sl Einstein\/}}
\newcommand{\tgt}{V603~Aql}
\newcommand{\gs}{GS~1843+009}
\newcommand{\mdot}{$\dot M$}
\newcommand{\cps}{ct\,s$^{-1}$}

\shorttitle{X-ray Observations of V603 Aql}
\shortauthors{Mukai \& Orio}

\begin{document}


\title{X-ray Observations of the Bright Old Nova V603 Aquilae}


\author{K. Mukai\altaffilmark{1}}
\affil{Code 662, NASA/Goddard Space Flight Center, Greenbelt, MD 20771}

\and

\author{M. Orio\altaffilmark{2}}
\affil{Istituto Nazionale di Astrofisica (INAF), Osservatorio Astronomico
	di Torino, Strada Osservatorio, 20, I-10025, pino Torinese (TO), Italy}


\altaffiltext{1}{Also Universities Space Research Association}
\altaffiltext{2}{Also Department of Astronomy, 475 N. Charter Str.,
	University of Wisconsin, Madison, WI 53706}


\begin{abstract}

We report on our \chandra\ and \xte\ observations of the bright
old nova, \tgt, performed in 2001 April, supplemented by our analysis
of archival X-ray data on this object.  We find that the \xte\ data
are contaminated by the Galactic Ridge X-ray emission.  After accounting
for this effect, we find a high level of aperiodic variability in the
\xte\ data, at a level consistent with the uncontaminated \chandra\ data.
The \chandra\ HETG spectrum clearly originates in a multi-temperature plasma.
We constrain the possible emission measure distribution of the plasma
through a combination of global and local fits.  The X-ray luminosity
and the spectral shape of \tgt\ resemble those of SS~Cyg in transition
between quiescence and outburst.  The fact that the X-ray flux variability
is only weakly energy dependent can be interpreted by supposing that
the variability is due to changes in the maximum temperature of the plasma.
The plasma density is likely to be high, and the emission region is likely
to be compact. Finally, the apparent overabundance of Ne is consistent with
\tgt\ being a young system.

\end{abstract}




\section{Introduction}

Cataclysmic variables (CVs) are interacting binary systems in which a
white dwarf primary accretes from a Roche-lobe filling, late-type secondary
(see \citealt{tome} for a comprehensive review).  In the majority of
CVs, the magnetic field of the primary is not strong enough to control
the accretion flow.  In such systems (non-magnetic CVs), accretion proceeds
via a Keplerian disk, which radiates half the available potential energy
mostly in the optical and UV.  The other half of the potential energy is
in principle available to power the boundary layer between the disk and the
white dwarf \citep{PR1985a,PR1985b}.  It is the likely source of X-rays that
are observed in nearby non-magnetic CVs, since the accretion disk is not
hot enough to be a significant source of X-rays.  There are, however,
important questions regarding the validity of this simple interpretation.  
X-ray observations are an essential tool that may help advance our
understanding of the accretion onto the white dwarfs in CVs in general,
and the nature of the boundary layer in particular.

A subset of CVs have been discovered as classical novae.  These eruptions
are due to thermonuclear runaway of hydrogen-rich material accreted on
to the primary of a CV.  \tgt\ was the brightest classical nova of the
20th century during its eruption (as Nova Aquilae 1918).  Classical novae
are spectacular laboratories of thermonuclear reactions, and make some
notable contributions to the chemical evolution of the interstellar medium.
However, the outburst characteristics, such as the energetics, the duration,
and nucleosynthesis yields, all depend on the details of the accretion
processes in quiescence, including their quiescent mass accretion rate
(\mdot).  This is an important motivating factor in attempting to learn
as much about the nova systems in quiescence.

Various authors have claimed \tgt\ to be magnetic; however, none of these
claims have been verified by later, more sensitive, observations.  The
claim of periodicity in polarization by \cite{HM1985} has been disproved
by more sensitive observations by \cite{C1986} and by \cite{Nea1996}.
Although there is a real modulation of the X-ray flux with a 60 minute
timescale in in the \einstein\ data \citep{USC1989}, this modulation is
neither persistent nor periodic, hence \tgt\ should be considered
a non-magnetic CV (\citealt{Bea2003}; \S 4.1).

\tgt\ is the brightest old (=quiescent) nova in the optical, UV and
X-ray wavelengths.  It is also relatively nearby; \cite{HD1927} have
estimated a distance $d$ of 360 pc using the expansion parallax method.
The statistical error is negligible in this estimate, although there
is a systematic error due to departure from spherical symmetry in nova
ejecta \citep{Wea2000}.  \cite{HG1988} estimate $d$ = 430 pc using the
correlation between the absolute magnitude and the rate of decline of
classical novae ($t_3$ = 8 days; \citealt{D1988}).  At such a distance,
Hipparcos parallax measurement does not lead to a more precise estimate
($d$ = 237$^{+380}_{-90}$ pc; \citealt{D1999}).  In this paper, we adopt
the \cite{HD1927} value of 360 pc as the fiducial distance.

\tgt\ is also a permanent superhumper showing both positive and negative
superhumps \citep{Pea1997,P1999}, suggesting a well-established
disk with high \mdot\ simultaneously undergoing apsidal and nodal precessions.
\cite{Fea1982} inferred a total disk luminosity of
$\sim$ 10$^{35}$ ergs\,s$^{-1}$ from UV data, requiring
\mdot\ $> 10^{18}$ g\,s$^{-1}$.  \tgt\ has been observed in X-rays with
\einstein, \exosat\, \rosat\ PSPC, \asca, and \xte.  \cite{Dea1983}
found the source to be 3 times brighter in \einstein\ IPC in their
observation ($\sim$ 0.7 ct\,s$^{-1}$, 1981 March) than in 1979 Oct
($\sim$ 0.3 ct\,s$^{-1}$). \cite{Dea1983} also inferred a very hard
(kT$>$20 keV) spectrum from IPC and MPC data, and a total flux of
L$_x= 3 \times 10^{33}$ egs\,s$^{-1}$, a conclusion that we now consider
suspect (\S 3.1).  Note that even this luminosity can be generated
by an accretion rate onto the white dwarf (\mdot) of
$\sim$ $10^{16}$ g\,s$^{-1}$, much lower than that inferred from
the accretion disk data.

\tgt\ is the second brightest non-magnetic CV observed with \asca\ 
\citep{Bea2004}, which makes it a prime candidate for detailed X-ray
spectroscopy.  In this paper, we present the result of our {\sl Chandra
X-ray Observatory\/} High Energy Transmission Grating (HETG) observation
of \tgt, with contemporaneous monitoring campaign using \xte.
We also briefly describe archival \asca\ and \xte\ data on \tgt.

\section{Observations}

\tgt\ was observed with \asca\ \citep{ASCA} during 1996 October 8--10
over a period of approximately 34 hrs.  After standard screening,
we have extaracted 64 s bin light curves in 2 bands ($<$2 keV and
$>$2 keV) as well as over the entire \asca\ energy range (0.5--10 keV),
combining data from all 4 instrument.  We have also
extracted 2 spectra, one for the 2 GISs combined and the other
for the 2 SISs combined, and associated response files.  This
observation is also featured by \cite{Bea2004}, who have analyzed
the complete set of \asca\ observations of non-magnetic CVs.

\tgt\ was observed 3 times with \xte\ \citep{RXTE}
during 1998 December 5--7, each over a period of 6--7 hours,
and over $\sim$52 hours of total elapsed time.
We have analyzed only the data obtained with the Proportional Counter
Array (PCA; \citealt{PCA}), which consists of 5 Proportional Counter
Units (PCUs).  We have applied the standard screening
criteria, after which we have accumulated 41 ksec of exposure.
The background was estimated using the faint source model using calibration
file dated 2003 March 30.

We have observed \tgt\ with \chandra\ \citep{CHANDRA} using the High Energy
Transmission Gratings (HETG; \citealt{HETG}) and with the ACIS-S CCD
array as detector \citep{ACIS} on 2001 April 19/20
over a span of 18 hrs.  Because \chandra\ is in a high orbit, there were
no interruptions during the observation, resulting in an exposure of 63.5 ksec.
We have used the pipeline-produced evt2 and pha2 files and used CIAO 3.0.2
with caldb 2.26 to generate the response file, which includes a calibration
of the contaminant on the ACIS detector.  We have combined $\pm$1st orders
of High Energy Grating (HEG) data into one spectral file, with the
response file appropriately combined; similarly for Medium Energy
Grating (MEG) data, with the $\pm$1st orders combined.  This observation
was previously included in a comparative study of \chandra\ HETG spectra
of 7 CVs \citep{Mea2003}.

Near the time of the \chandra\ observation, we have monitored \tgt\ with
\xte.  We have obtained twenty short observations over 2001 April 17
through 22, each resulting in a typical exposure time of 2--3 ksec,
for a total of 55.4 ksec of good time.  For these as well as for the
1998 \xte\ observations, light curves are presented in the form
of cts\,s$^{-1}$ per PCU, since variable numbers of PCUs were active
at any given time.  In addition, for the 2001 observations, we
have excluded the data taken with PCU0, as the propane leak in this
detector makes background subtraction problematical\footnote{See
http://rxte.gsfc.nasa.gov/docs/xte/oldnews/pcu\_0.html.}.

We present summaries of these observations in Table\,\ref{obstab}.
Earlier observations of \tgt\ with \einstein, \exosat, and \rosat\ are
summarized in \citet{Oea2001}.

\section{Analysis and Results}

\subsection{Contamination in Non-imaging X-ray Data}

\tgt\ is located in a crowded region near the Galactic plane
(at $l=33.16^\circ$, $b=0.83^\circ$).  Background subtractions
for \xte\ PCA data are performed using a model that accounts for
the time-variable particle background and for the average, high-latitude,
cosmic X-ray background.  Galactic diffuse emission
remains in the ``background subtracted'' data, as well as any specific
X-ray sources that are in the field-of-view.  The latter includes the
supernova remnant Kes 79 \citep{Sea2003}, 53.8$'$ away, consisting
of a thermal shell and a soft, central point source.  This source is
probably too soft and too distant to have a significant impact.
More worrisome is the recurrent transient X-ray pulsar, \gs\ (\citet{Iea2001}
and references therein), located 52.2$'$ away from \tgt.  During an
outburst, this pulsar can be much brighter than \tgt, particularly
at high energies since \gs\ is a hard source, and can strongly contribute
to the counts of non-imaging observations aimed at \tgt.

It is quite possible that the \einstein\ MPC data, taken in 1981 March
and is reported by \citet{Dea1983}, were strongly contaminated by \gs,
which was unknown at the time (its discovery was in 1988; \citealt{Kea1990}).
Note that the MPC has a relatively wide (1.5$^\circ$ FWHM) filed-of-view
and there are no contemporary all-sky monitor data. The high
apparent ratio of MPC to IPC count rates (3.24/0.71) and the
hard spectral shape inferred from the MPC data are both incompatible
with our findings below.  Either \tgt\ was in a different state
in 1981 March, or the MPC data were contaminated by a hard source,
such as \gs.  We have extracted the MPC data from HEASARC and searched
for the 29.5 s spin period of \gs; although we do not see the period,
this unfortunately still allows the possibility of a significant
contamination by this pulsar, since the pulse fraction of \gs\ is
unusually small (4\% peak-to-peak; \citealt{Kea1990}).

\xte\ PCA is less susceptible to contamination by this source, because
of its narrower field-of-view (1$^\circ$ FWHM).  Moreover, we have checked
\xte\ ASM data to ensure that, at the time of these \xte\ observations,
\gs\ was not in an active state.  Although we cannot exclude contamination
at a low level by \gs\ or another source, it probably is not a major concern.

However, the PCA data are contaminated by the apparently diffuse,
``Galactic Ridge'' X-ray emission \citep{Wea1985,VM1998}.  From
Figure 1 of \citet{Wea1985}, we estimate a 2--6 keV flux of the Ridge
of order 7$\times 10^{-12}$ ergs\,s$^{-1}$cm$^{-2}$ per \exosat\ ME
beam (45$'$ FWHM) near \tgt.  Taking into account the larger beam size
of \xte\ PCA (1$^\circ$ FWHM), we expect of order 2 cts\,s$^{-1}$ per
PCU from the Ridge in the observations of \tgt\footnote{Similarly,
the Ridge definitely contributed significantly to the observed
\einstein\ MPC count rate.  However, the Ridge spectrum is quite similar
to that of \tgt, so a contribution from \gs\ remains a likely possibility.}.
Although the average spectrum of the Ridge has been precisely characterized,
the fact that there are at least two components whose relative importance
is a function of Galactic coordinates \citep{VM1998} makes it difficult
to subtract the Ridge contamination with confidence.  In the rest of this
work, we will use the \xte\ data primarily to discuss the short timescale,
relative variability of \tgt.

\subsection{Variability}

Despite the above considerations, the long-term variability of \tgt\ is
well established through imaging observations \citep{Oea2001},
for example by the different count rates recorded by the \rosat\ PSPC.
Here we investigate short-term ($\sim$5 days or less) variabilites
using \asca, \xte, and \chandra\ data.

\begin{figure}
\plotone{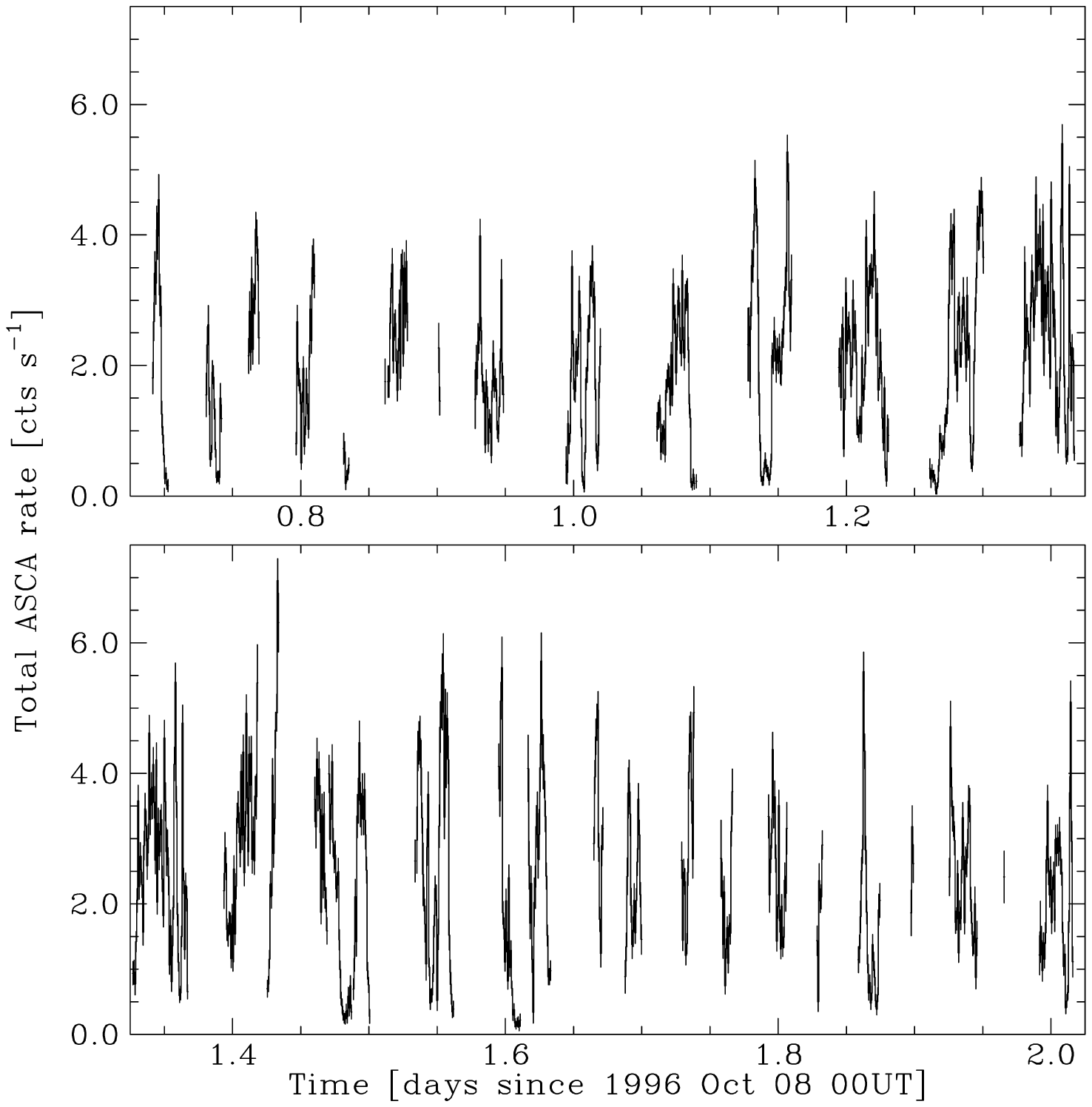}
\caption{\asca\ light curve of all instruments combined and for
all energies, in 64 s bins.  Time increases from left to right in
the top panel, then continues in the bottom panel.  One \asca\ orbit's
worth of data are plotted on both panels for continuity.}
\label{ascalc}
\end{figure}

\begin{figure}
\plotone{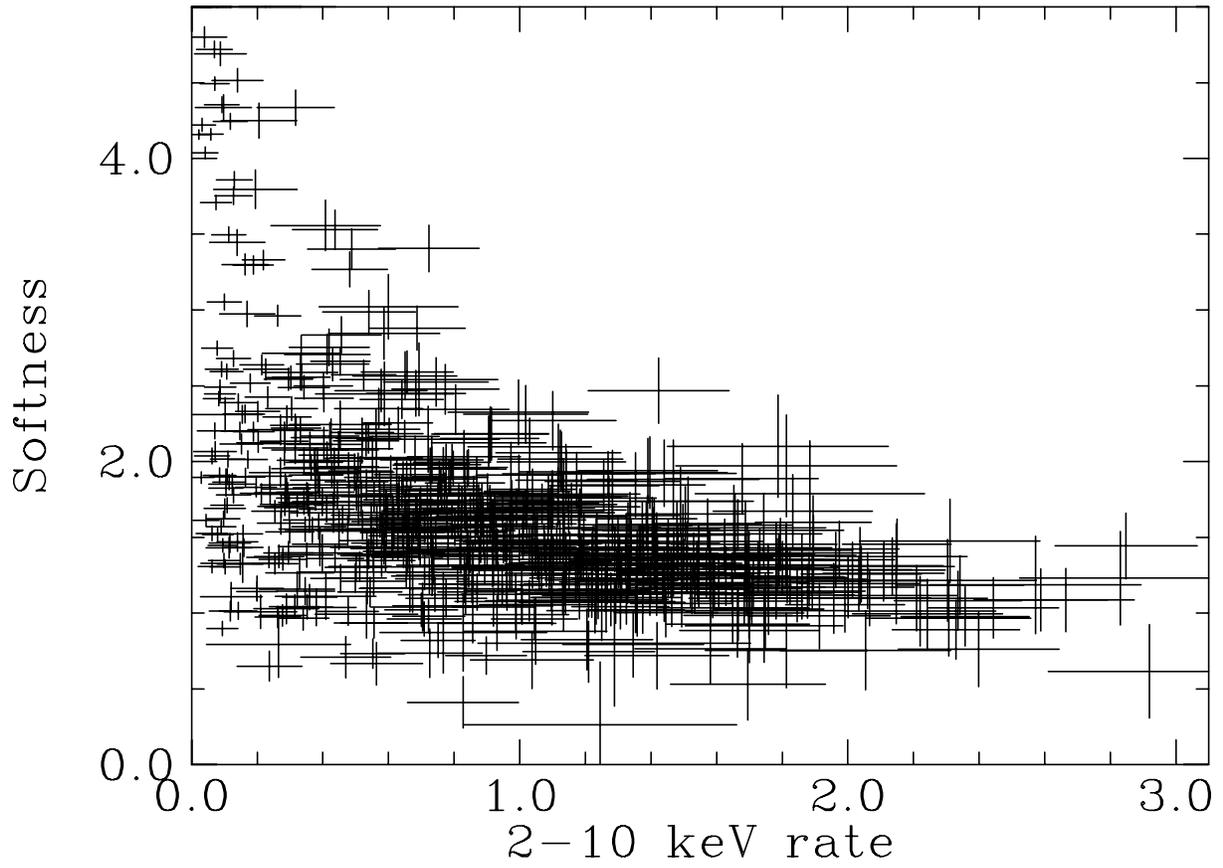}
\caption{Softness ratio ([$<$2 keV count rate]/[2--10 keV count rate])
plotted against the 2--10 keV count rate, from the 64 s bin \asca\ light
curves.}
\label{ascasoft}
\end{figure}

We present the 64 s bin \asca\ light (0.5--10 keV) in Figure\,\ref{ascalc}.
A significant variability is obvious: although the average count rate is
2.22 \cps, values ranging from $<$0.1 to $\sim$7.0 \cps\ (64 s bins, all 4
instruments and all photon energies combined) are seen.  Despite the strong
variability, there is no single stable period in this data \citep{Bea2004}.
We have also investigated the energy dependence of this variability.
In Figure\,\ref{ascasoft}, we plot the softness ratio (counts below
2 keV divided the 2--10 keV counts) against the 2--10 keV count rate.
Except at the lowest count rates, this plot is relatively flat,
and shows a slight softening trend when the count rate is lower.
This rules out changes in photoelectric absorption, which would predict
a strong hardening at lower count rates, as a major cause of variability.

\begin{figure}
\plotone{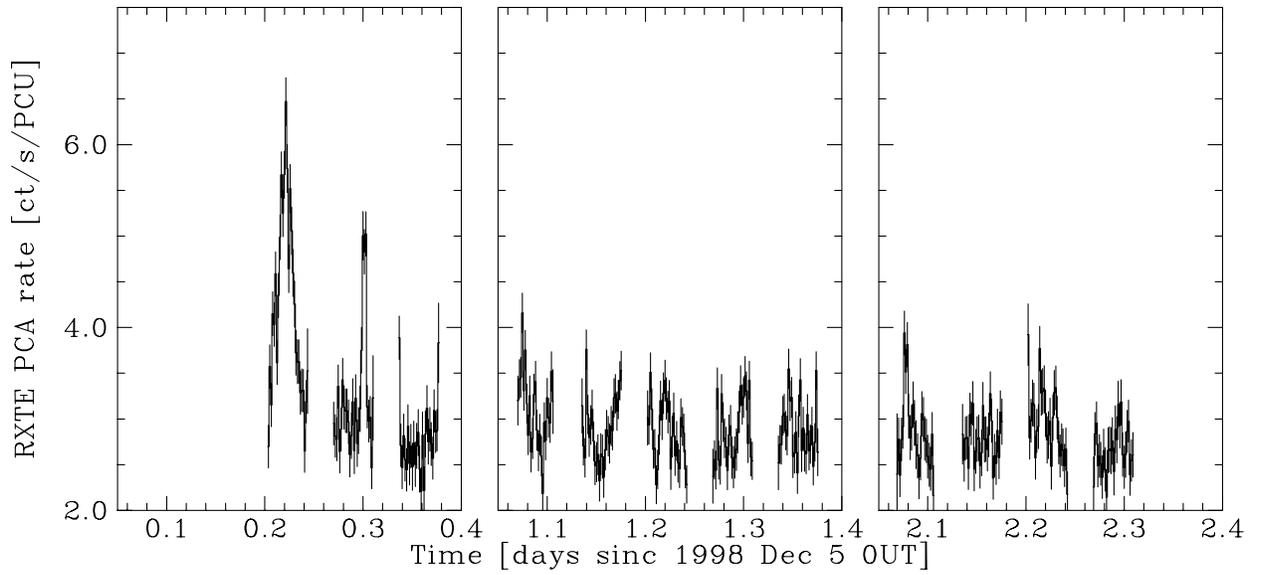}
\caption{\xte\ PCA light curves of in 16 s bins from the 1998
observations.  Particle and high Galactic latitude background have
been subtracted, but not the Galactic background.  The latter is
estimated to be approximately 2 \cps, which has been chosen as the
origin of Y axis of this figure.}
\label{xte1998lc}
\end{figure}

We present the light curve of the 1998 \xte\ observations in
Figure\,\ref{xte1998lc}.  As the light curve contains counts
due to the Galactic Ridge, we have chosen 2.0 (the
estimated Ridge contribution) as the origin of the Y axis.
With this offset, the high level of variability is apparent,
including a large flare-like event at the beginning of the
first observation.

\begin{figure}
\plotone{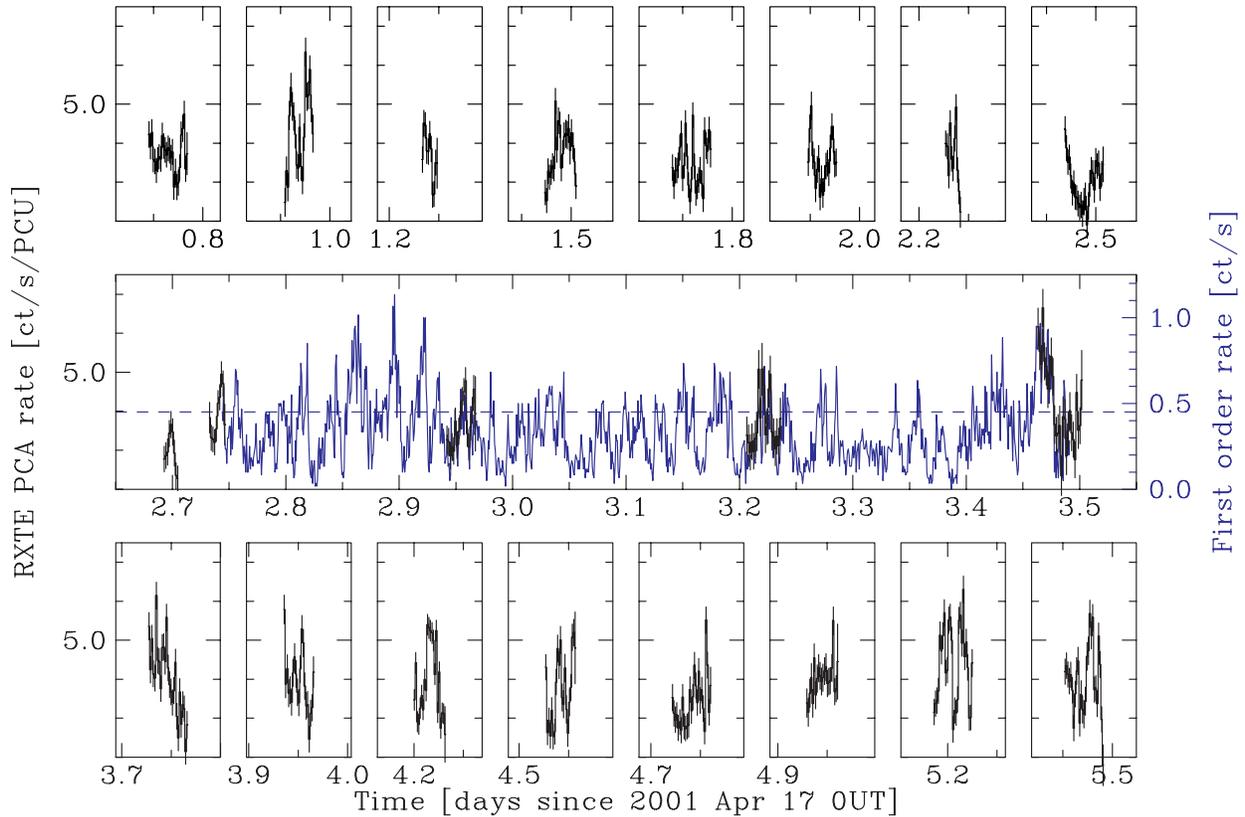}
\caption{\xte\ PCA light curves in 16 s bins, and \chandra\ HETG
1st order light curves in 60 s bins, of \tgt\ from the 2001 observations.
The middle panel shows the HETG curve in blue (with scales labelled
on the right) and the PCA curves from 4 overlapping pointings in black.
The horizontal dashed line is the HETG count rate used to divide the
data into high and low states for Figure\,\ref{rplot}.
The top and bottom panels show the non-overlapping PCA curves;
all PCA light curves are offset by 2 \cps.}
\label{apr2001lc}
\end{figure}

Although not the main strength, the \chandra\ HETG/ACIS-S observation
provides an uninterrupted light curve of \tgt.  Because the zeroth order
(undispersed) image is heavily affected by pile-up, we restrict our
analysis to the first order photons.  We have combined the HETG 1st
order light curve and the 2001 PCA light curves into a single figure
(Figure\,\ref{apr2001lc}).  Of the 20 separate \xte\ pointings,
4 overlapped in time with the \chandra\ observation.
The middle panel shows the light curves of \tgt\ during this
interval, both using the HETG (blue) and the PCA (black),
the latter plotted with the $\sim$2 \cps\ per PCU offset as before.
These two light curves overlap closely.  This suggests
that our estimate of the Galactic Ridge contribution to the PCA
data is a reasonable one.  Moreover, any energy dependence of the
variability must be small, since the two instruments are sensitive
to different energy ranges.

\begin{figure}
\plotone{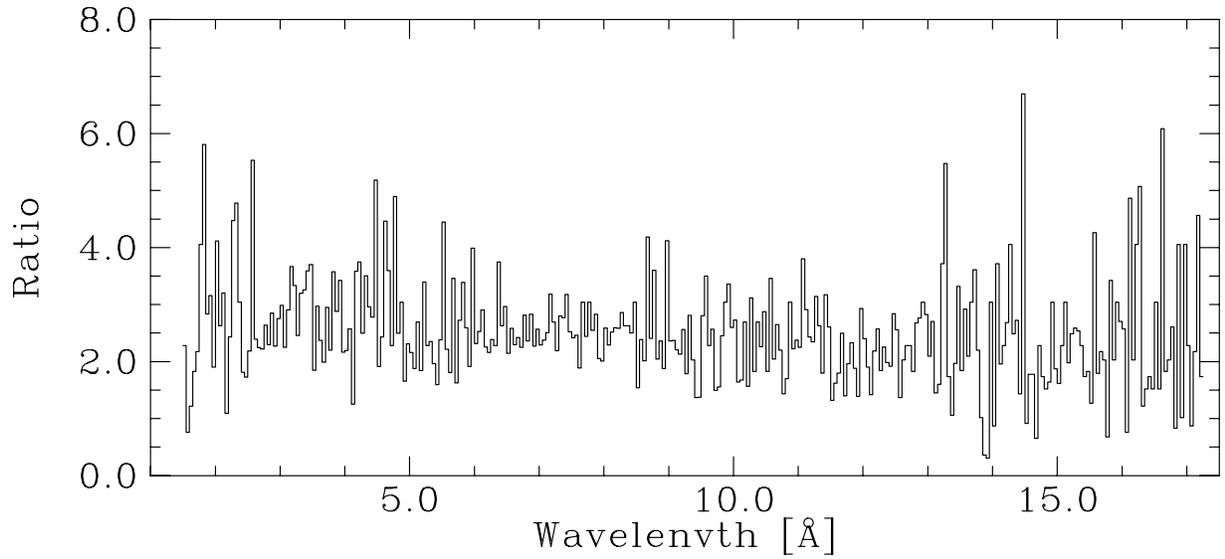}
\caption{Ratio of the \chandra\ HETG (MEG) high state spectrum to
the low state spectrum.}
\label{rplot}
\end{figure}

To investigate the energy dependence of the variability further,
we have extracted spectra from the \chandra\ HETG (MEG) data
from the high and low states, defined as when the total
1st order count rate was higher than/lower than 0.45 \cps\ (as
shown by the dashed line in the middle panel of Figure\,\ref{apr2001lc}).
This ratio is shown as a function of the wavelength in
Figure\,\ref{rplot} in the 1.5--17.2\AA\ (0.72--8 keV) range.
We confirm the \asca\ result (Figure\,\ref{ascasoft}) of
mild energy dependence.  With this definition of high and low
states, the high state spectrum is somewhat harder: a linear
fit to the trend suggests high/low ratio is $\sim$20\% higher
at 8 keV than at 1 keV.  We reach a similar conclusion from the
energy independence of the variability in the \xte\ PCA data
(without subtracting the Galactic Ridge, whose precise spectrum
at the location of \tgt\ is uncertain).  This suggests again that
the variability of \tgt\ is only mildly energy dependent, and also
that the spectra of \tgt\ and of the Galactic Ridge are similar,
barring an unfortunate coincidence.

\begin{figure}
\plotone{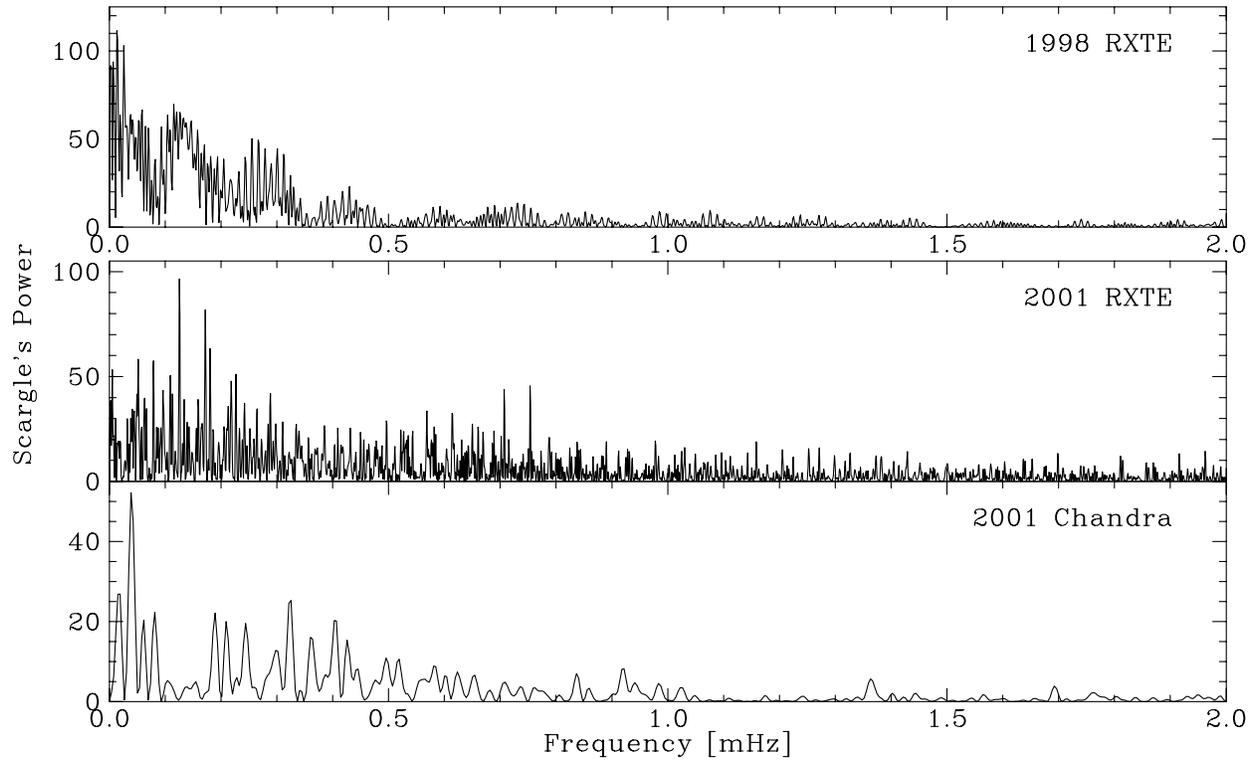}
\caption{Power spectra of \xte\ PCA and \chandra\ HETG data
plotted against frequency in mHz.}
\label{sgmplot}
\end{figure}

We present the power spectra (using the definition of \citealt{S1982})
from the 1998 and 2001 observations in Figure\,\ref{sgmplot}.
\citet{Bea2004} contains a corresponding power spectrum plot from
the \asca\ observation.  We confirm their finding that there is no
coherent periodicity in the X-ray data of \tgt. In particular, the
power spectrum of the \chandra\ data does not suffer from the aliasing
problem that plagues low Earth orbit satellite data, and thus the
multiple peaks seen in the bottom panel between 0.2--0.5 mHz
(periods in the 2000--5000 s range) must be taken at face value.
That is, \tgt\ shows strong variability on these timescales,
which however is not due to a single, coherent period.

In addition, we have attempted to verify the tentative detection
of power at 60 hrs \citep{Oea2001} using the 2001 \xte\ campaign,
but failed to do so.  This is in part due to the the combination
of the powers at higher frequencies (0.2--0.5 mHz) and our sampling
pattern.  We also note that the earlier claim was based on \rosat\ data,
which leaves open the possibility of a soft X-ray absorption event
that would be undetectable with \xte.  Finally, the highest peak in
the \chandra\ power spectrum corresponds to a period of ~25,000 s.
This is a significant fraction of the duration of our \chandra\ data,
so we caution against overinterpreting this peak.

\subsection{Spectroscopy}

\begin{figure}
\plotone{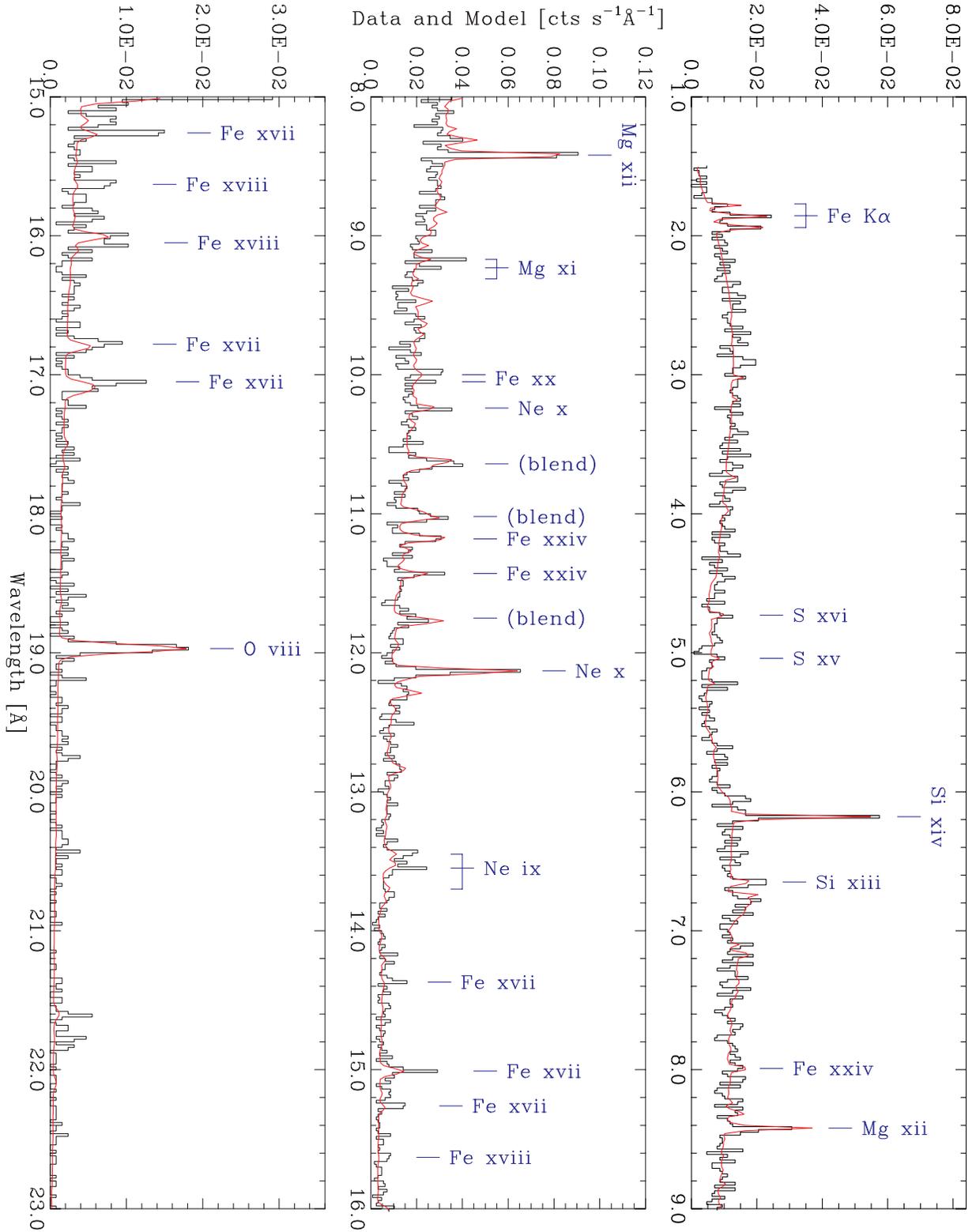}
\caption{The average spectrum of \tgt\ as observed with \chandra\ HETG/ACIS-S,
plotted in 3 sections.  For clarity, we plot only the HEG data in the
top panel, and only the MEG data in the bottom two panels.  The data
are shown as black histograms, while the best fit cooling flow model
(see text for details) are shown as red lines.  Emission lines that
are significantly detected are labeled.}
\label{hetgov}
\end{figure}

We show the average \chandra\ HETG/ACIS spectrum of \tgt\
in Figure\,\ref{hetgov}.  The line-rich nature of its spectrum
is obvious.  The He-like and cold Fe K$\alpha$ lines are strongly
detected, as well as a weaker H-like Fe K$\alpha$.  The H-like lines
of Si, Mg, Ne and O are prominent, while their He-like counterparts
are detected but significantly weaker.  In addition, many Fe L-shell
lines are detected in the 10--17 \AA\ range.
This overall appearance is that of an X-ray emission from multi-temperature,
collisionally excited plasma, as was noted earlier by \citet{Mea2003}.

We have analyzed this spectrum using two complementary methods.
In one, we have characterized individual lines through local fits,
using an unabsorbed power law continuum and Gaussian lines.  In the
second method, we have performed a global fit using the variable
abundance version of the cooling flow model ({\tt vmkcfl})
with the Gaussian broadening function ({\tt gsmooth}), as well as
an additional Gaussian representing the fluorescent Fe line at 6.4 keV.

In principle, we can derive several quantities from the global fit.
Chief among them are the maximum temperature of the plasma,
the global mass accretion rate, and the various elemental abundances.
We do not detect a significant interstellar absorption; in fact, our
upper limit ($1.3 \times 10^{19}$ cm$^{-2}$) is considerably smaller
than the value of 6--8$\times 10^{20}$ cm$^{-2}$ inferred from
\rosat\ PSPC data by \citet{Oea2001}.  This may indicate
a possible problem with the calibration of contaminant for
\chandra\ HETG/ACIS-S data, which would be a source of systematic
errors in the low energy results, both from global and local fits.
We also obtain only an upper limit for the minimum temperature,
$kT_{\rm min}$. The fact that the model is pegged at its hard lower
limit (0.0808 keV) suggests the limitation in the grid of models,
in addition to the lack of sensitivity of HETG to such low temperature
components, may be playing a role.  The maximum temperature is found
to be 20.3$^{+1.1}_{-1.8}$ keV.  As for the abundance values, we
only report our results for Ne and Fe, two elements that are clearly
non-solar relative to the fiducial values of \cite{AG1989}.  Other
abundances are consistent with solar, although the best-fit values may
deviate from 1.0 by 0.1 or more.  The accretion rate of the plasma
that is emitting the observed X-rays is determined to be 
8.00$\pm 0.01 \times 10^{-11}$ M$_\odot$\,yr$^{-1}$ for the assumed
distance of 360 pc.  We present a summary of the global fit results
in Table\,\ref{gfres}, with formal 90\% confidence errors, and further
consider the reliability and the implications of this procedure
in \S 4.2 below.

We list all the lines that are confidently detected in Table\,\ref{llist}.
For each line, we list the probable identification, observed wavelength,
the velocity width, and the photon flux according to the local fit.
These lines are also labelled in Figure\,\ref{hetgov}.  The measured
flux of a line can be combined with the theoretical curve of
its emissivity in a collisionally excited plasma as a function of
plasma temperature to draw a curve on the temperature -- emission
measure plane (Figure\,\ref{emplot}).  For Ne and Fe, we have used
the best-fit abundance values from the global fit in placing these
curves.  Any point on one of these curves represents the maximum
emission measure a plasma of that temperature can have, before
predicting more flux in that line than is observed.  For multi-temperature
plasma, emission measures must be summed at several discrete temperatures
or integrated over the relevant range.  This is analogous to the differential
emission measure (DEM) distribution analysis favored for the study of
stellar coronae (see, e.g., \citealt{GJ1998}).   Note, however, we have
not derived the true DEM distribution unambiguously. What we have done
here is to constrain the shape of the DEM distribution from local fits,
and to fit a discrete approximation of the DEM distribution appropriate
for a isobaric cooling flow.  Note also that our data for \tgt\ mixes
lines from different elements, and lines of different types, and this
complicates the interpretation.  Nevertheless, we find it satisfactory
that the line from the global fit parallels the locus defined by the
series of parabola-like curves derived from individual line fluxes,
with an appropriate offset (a factor of 4, which is roughly the
number of temperature steps of the global model within the typical width
of the parabola-like curves).

It is also possible to study selected lines in greater detail.
In Figure\,\ref{fek}, we present a close-up view of the Fe K$\alpha$
region of the spectrum as observed with the HEG.  The equivalent width
of the fluorescent (6.4 keV, or 1.98\AA) line is 150 eV, which is what
is expected of reflection off a surface that subtends a 2$\pi$ steradian
as seen from the primary X-ray source, if the Fe abundance is roughly solar.
In Figure \,\ref{triplets}, we present close-up views of the He-like
triplet regions for Mg and Ne.  In both cases, we detect lines near
the expected wavelength of the forbidden component.  However, there
appears to be an offset.  Given this, the possibility that these
apparent detections are due to interlopers should be kept in mind.
We place an upper limit to the flux of the forbidden component at
laboratory wavelength of 2.8$\times 10^{-6}$ photons\,cm$^{-2}$s$^{-1}$
for Mg\,XI and 3.0$\times 10^{-6}$ photons\,cm$^{-2}$s$^{-1}$ for
Ne\,IX, respectively.  We explore the possible implications further
in \S 4.4 below.

The line width is generally small.  Most lines are consistent
with a velocity width of order 200--300 km\,s$^{-1}$.  This is not
surprising given the low inclination angle of \tgt\ ($i \sim 20^\circ$;
\citealt{Pea1993}).  The exceptions are the Fe fluorescent line and
several observed lines that are likely to be blends.

Finally, we confirm that the same spectral model can be used to fit the
\asca\ spectra (those taken with GIS and with SIS) of \tgt.  We obtain
an acceptable fit just by adjusting the N$_{\rm H}$ and the participating
mass accretion rate.  The former is likely due to calibration issues
(both for \asca\ and for \chandra), while the latter is not surprising,
given the high level of variability known in this system.  We obtain
a participating mass accretion rate of $\sim 1.6 \times 10^{-10}$ 
M$_\odot$ yr$^{-1}$ during the \asca\ observation.  We can also fit
the Ridge-contaminated \xte\ data with the same model, again suggesting
the close similarity between the spectra of \tgt\ and the Ridge,
but we refrain from further interpretation due to this contamination
problem.

\section{Discussion}

\subsection{Confirmation of the non-magnetic nature of \tgt}

Our analysis of X-ray curves fails to show a persistent periodicity.
Although the power spectra of individual observations often show
prominent peaks, there is no single period that appears in
all observations.  Instead, the power spectrum of our \chandra\ data
show the simultaneous presence of many peaks.  Moreover, the nature
of X-ray variability in \tgt\ is atypical for an IP.  If there is an energy
dependence in the X-ray variability of \tgt, it is rather subtle;
a strong energy dependence, on the other hand, is a common characteristic
of IPs.

Combined with other recent X-ray results \citep{Bea2003,Bea2004},
as well as non-detection of polarization \citep{C1986,Nea1996},
we affirm the conclusion that \tgt\ is a non-magnetic CV.  The earlier results
are likely correct in finding power at certain frequencies in the
X-ray data, but are incorrect in interpreting the power spectrum
as evidence of a single, coherent, underlying clock.  We consider
the nature of the hard X-ray variability further in \S 4.3 below.

\subsection{Interpreting the Global Fit Results}

Scientifically, the fact that we must assume a physical model
for the X-ray emission in a global fitting approach is both its strength
and its weakness.  It allows physical interpretations of the
fit results, on the one hand.  On the other hand, the interpretations
are only as good as the set of assumptions one has adopted.
In our global fits, we have assumed a isobaric cooling flow
with no additional heating or cooling mechanisms, absorbed only by
the interstellar medium.  Several authors have already explored
alternative assumptions while analyzing the \chandra\ HETG data of other
CVs \citep{Pea2003,Hea2004}.  Moreover, we cannot rule out the more simple
model in which the emission measure is a power law function of temperature
({\tt cevmkl} model in xspec; see, for example, \citealt{Bea2004}).  As
Figure\,\ref{emplot} shows, the isobaric cooling flow model deviates only
mildly from a power law above temperature of 3$\times 10^6$K, where the
HETG data have good sensitivity.  In this respect, we note that our
best-fit global model fails to reproduce the observed Fe\,XVII line fluxes
(Figure\,\ref{hetgov}), and that a {\tt cevmkl} model with a reduced O
abundance is a viable model.  Nevertheless, we discuss below
our global fits using the isobaric cooling flow model, since it appears
adequately to describe our data and has a physical interpretation.

The numerical complexity and the degeneracy of model parameters
are further factors limiting what one can achieve with global
fits.  For example, the maximum temperature (kT$_{\rm max}$) can
only be constrained if the overall abundance is fixed at an assumed
value.  This is because, for the likely range of kT$_{\rm max}$
($>$10 keV), the hottest plasma contributes little line fluxes, and the
continuum shape within the \chandra\ HETG band changes only moderately.
The main effect of increasing kT$_{\rm max}$ is to reduce the
equivalent widths of the lines, which can also result from
lower abundances.  We have therefore fixed the abundances
to the solar values of \cite{AG1989}, with the exception of
Ne and Fe.  Moreover, we have not included reflection in our
fits, since \chandra\ HETG has little sensitivity to reflected
continuum, even though the strength of the 6.4 keV line indicates
that reflection must be occurring in \tgt.  Given these assumptions,
kT$_{\rm max}$ is tightly constrained.  Unlike in magnetic CVs,
however, there is no obvious and simple relationship linking
kT$_{\rm max}$ with the white dwarf mass; this would require
a knowledge of exactly how the plasma is shocked in an accretion
disk boundary layer.

The total mass accretion rate that participate in the cooling flow
is well constrained, for the same solar abundance assumption,
and for the assumed distance.  However, as we discuss in \S4.4,
some of the lines have the potential as plasma diagnostics;
the other side of the coin is that the cooling flow model fit,
which does not account for effects of line opacity, photoexcitation
etc., may not be completely reliable.  Moreover, the lines may have
variable widths and/or profiles, which could skew our fit results:
physically, we expect the plasma
to continue to decelerate as it cools, so the line width should
be a function of the typical temperature needed to emit that line.
We approximate this using the {\tt xspec} convolution model, {\tt gsmooth},
which allows the line width to be a function of the photon energy
of that line.  These obviously are not the same thing.  However,
the unconfused lines are marginally resolved at best in our data
(see Table\,\ref{llist}), so this is probably not a major concern.

With these caveats, the mass accretion rate in the cooling flow is
tightly constrained in our global fit ($8.0 \times 10^{-11}$
M$_\odot$\,yr$^{-1}$ or $5.1 \times 10^{15}$ g\,s$^{-1}$).  In
addition, the observed luminosity in the 1.0--7.0 keV band is
1.88$\times 10^{32}$ ergs\,s$^{-1}$, depending only on the
assumed $d$, and hence is more robust. That is, \tgt\ is hard X-ray
luminous for a non-magnetic CV.  In particular, \tgt\ is significantly
more luminous than the dwarf nova Z~Cam in outburst
(6$\times 10^{30}$ ergs\,s$^{-1}$; \citealt{Bea2001}) or the inferred hard
X-ray component of the nova-like system, UX~UMa
(1.26$\times 10^{31}$ ergs\,s$^{-1}$; \citealt{Pea2004}).

SS~Cyg has a 3--20 keV luminosity of 3--9 $\times 10^{31}$ ergs\,s$^{-1}$
during the peak of normal outbursts \citep{Wea2003,Mea2004}, again fainter
than \tgt.  SS Cyg, however, has been observed at X-ray luminosity comparable
to that of \tgt: in the 1993 May \asca\ observation obtained towards the end
of an anomalous outburst \citep{Nea1994,Bea2004} and transitions into and out
of outburst \citep{Wea2003,Mea2004}.  Thus, SS Cyg has three main
X-ray/optical states.  In the presumed order of increasing mass
accretion rate onto the white dwarf, these are: the quiescence,
the transition, and the outburst peak.  Of these three, \tgt\ resembles
the transition most closely, in terms of its X-ray luminosity
and its X-ray spectrum.  The lack of detection of soft X-rays
(i.e., the high energy end of the expected EUV/soft X-ray blackbody like
component that would escape interstellar absorption)
in \tgt\ \citep{Oea2001} is also consistent with this
analogy. Arguing against this is the high optical/UV luminosity
(10$^{35}$ erg\,s$^{-1}$; \citealt{Fea1982}) of its accretion disk.

The observed X-ray luminosity is less than one per cent of
the observed UV luminosity.
If the UV luminosity of \tgt\ indeed represents the accretion
luminosity of its disk, the resemblance to the transition in
SS~Cyg must be a coincidence, and over 99\% of the expected luminosity
of the boundary layer remains unobserved.  If this luminosity
is radiated from the white dwarf surface, this must cover a large fraction
of the white dwarf surface to have escaped detection: 10$^{35}$ erg\,s$^{-1}$
radiated from 5\% of the area of a 6,000 km radius white dwarf would
have an effective temperature $kT$ of 25 eV, and would have been detectable
as a bright, soft X-ray component, even with a column of
N$_{\rm H} = 8 \times 10^{20}$ cm$^{-2}$.  At $kT$=20 eV, this is
not necessarily the case (most of the emission is in the EUV and
can be hidden by the interstellar absorption),
but the area must be 15\% of the white dwarf surface.
Thus, the standard picture of \citet{PR1985b} with an optically
thick boundary layer producing EUV emission, with an optically thin
hard X-ray emission near its surface, is a valid model for \tgt\ as
long as detailed models allow such a boundary layer to cover a large
fraction of the white dwarf surface.

We believe it is appropriate to question if all the observed
UV luminosity is directly from viscous heating in the accretion
disk.  \citet{Bea2003} note that the apparent correlation (with little
or no time lag) between the X-ray and UV fluxes may indicate that
a fraction of the UV flux is due to reprocessing of X-ray photons.
For this to be the case, we must postulate a luminous EUV component
that varies in phase with the observed hard X-rays, because the
luminosity of the latter is clearly insufficient to power the UV
component.  The hot white dwarf may also contribute non-negligible
EUV flux in this regard.  Our understanding of when and how nuclear
burning terminates in classical novae is still rather vague
(see, e.g., \cite{Oea2001}), and little is known about the
post-nova cooling.   If the entire photosphere of the white dwarf
in \tgt\ is hot, though less so than in super-soft source stage
of classical novae, its photospheric emission can be hidden more
easily than a soft X-ray component from a boundary layer.  The difficulty
with this picture is that one would expect the photospheric EUV luminosity
of a white dwarf to be constant, which is a significant difficulty in
explaining the strong UV variability\footnote{On the other hand,
we also expected the photospheric supersoft emission in immediate
post-novae to be constant, which have been proved incorrect in two
recent novae, V1494~Aql \citep{Dea2003} and V4743~Sgr \citep{Nea2003}.}.

If the correlated X-ray/UV variability in \tgt\ is confirmed,
then the accretion rate inferred from the UV luminosity may be
an overestimate, regardless of what we might eventually conclude
about the nature of the irradiating source.  In any case, the
resemblance between the X-ray behaviors of \tgt\ and SS~Cyg in
outburst transition requires an explanation.  Further observations
of the overall energy balance of SS~Cyg during transition may prove
useful in this regard.

\subsection{The nature of the short time scale variability}

We have argued earlier that the rapid X-ray variability of \tgt\ is
not due to photoelectric absorption.  There is, however, one parameter
in the cooling flow model that can explain the type of variability
observed: kT$_{\rm max}$.  When it is higher, the continuum is stronger
and somewhat harder; when lower, weaker and somewhat softer\footnote{This
does not contradict the earlier statement that the continuum shape
does not strongly constrain kT$_{\rm max}$.  The continuum shape can
only constrain kT$_{\rm max}$ to within $\sim$10 keV, while the line
to continuum ratio can constrain it to better than 1 keV, if the overall
abundances are fixed.}.  In magnetic CVs, such changes would be unphysical,
since the accretion flow is radial and kT$_{\rm max}$ is strongly tied
to the white dwarf mass.  In a non-magnetic CV, however, the shock
geometry is complex and three dimensional, and we know of no reason why
the shock geometry, hence kT$_{\rm max}$, cannot vary with a timescale
of 10s of minutes.  On the other hand, this timescale is too long
to be intrinsic to the boundary layer; some external mechanism to
drive such a change is probably necessary.  A search for similar
X-ray variability in other non-magnetic CVs, such as SS~Cyg during
transition, would be worthwhile.

This timescale, however, is similar to that seen in the UV
spectroscopy of \tgt\ \citep{Pea2000}.  Based on the detection
of episodic UV absorption lines, they suggested the presence of
a time-variable wind in \tgt.  This may be another symptom of
the underlying mechanism that causes the X-ray variability.
This is consistent with the fact that UV flux appears to be
correlated with X-ray intensities \citep{Bea2003}.

\subsection{Plasma Diagnostics}

\begin{figure}
\plotone{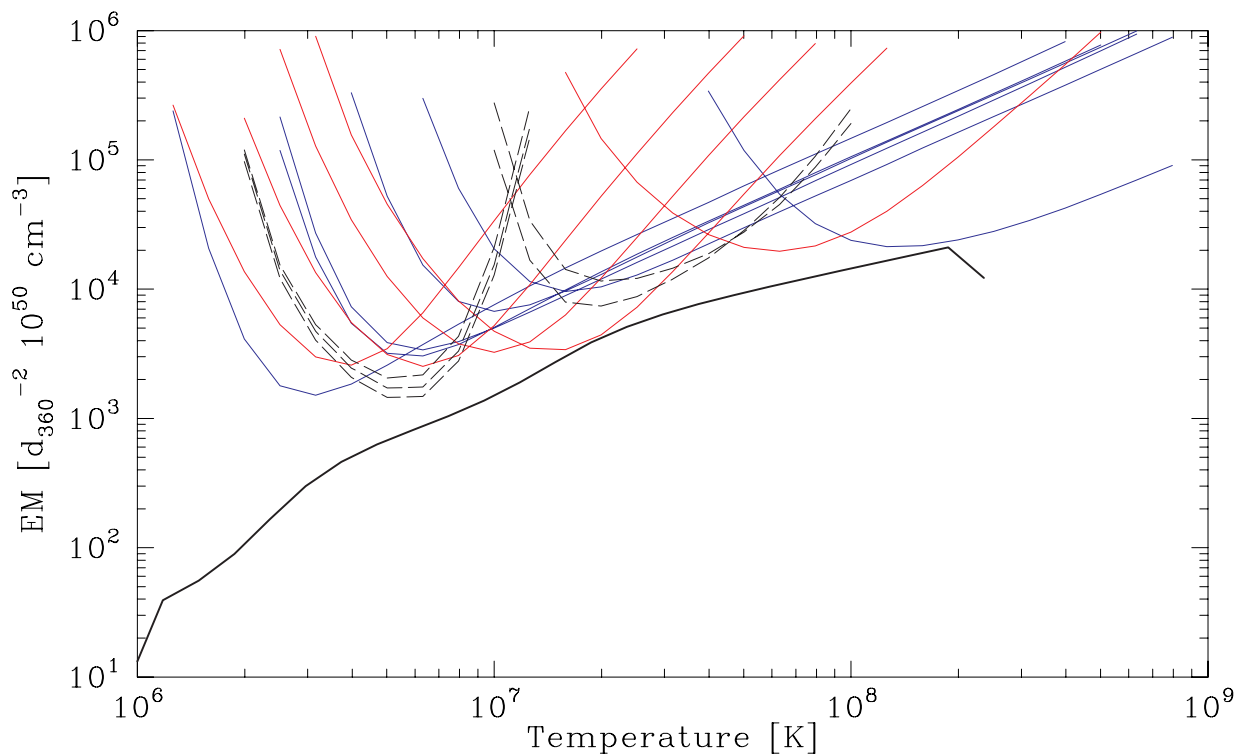}
\caption{The emission measure distribution as derived from the global fit,
and the constraints on it derived from individual line fits.  The latter
curves are shifted down by a factor of four, which approximately accounts
for the width of these parabola-like curves compared with the sampling
of the emission measure distribution from the global fit. Blue curves are
for H-like lines, red for He-like lines; black dashed lines are for
Fe L shell lines.}
\label{emplot}
\end{figure}

\begin{figure}
\plotone{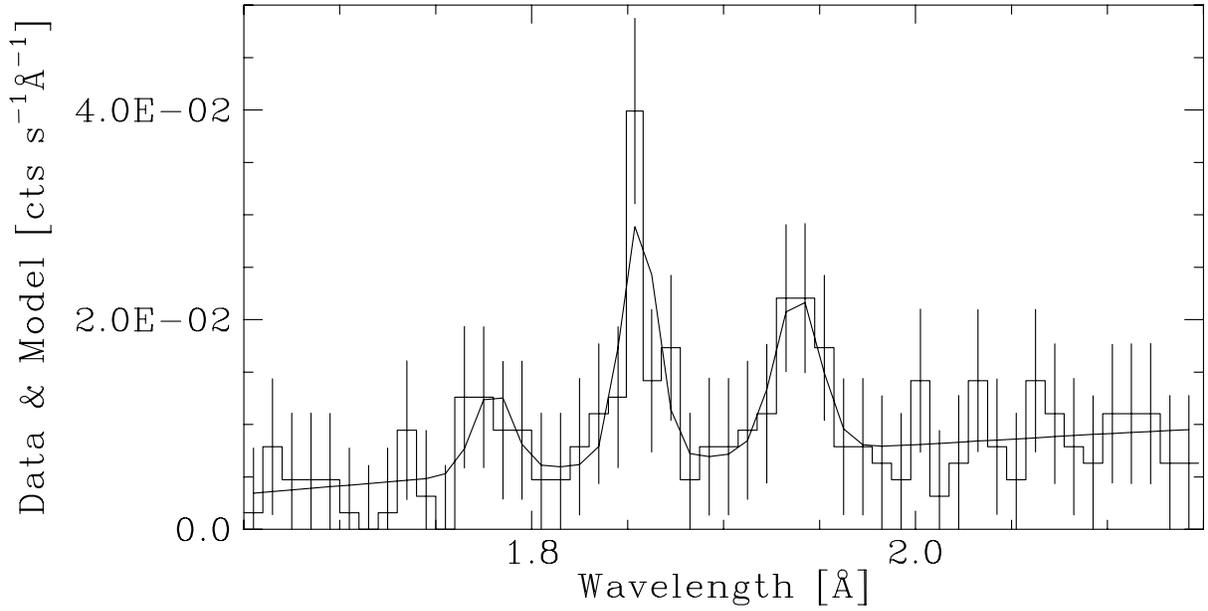}
\caption{The details around the Fe K$\alpha$ line; the HEG data and
the best local fit model are shown.}
\label{fek}
\end{figure}

\begin{figure}
\plotone{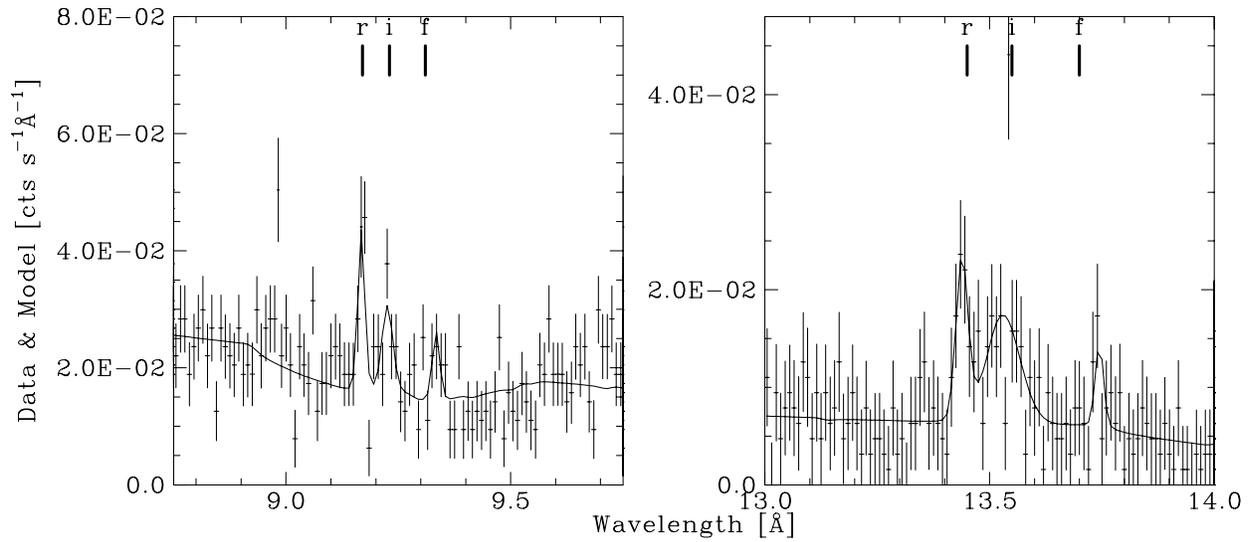}
\caption{The details around Mg and Ne helium-like triplets;
the MEG data and the best local fit models are shown.}
\label{triplets}
\end{figure}

The detection of individual lines opens up the possibility to
use selected line ratios as diagnostics of the temperature
and the density of the emitting plasma.  The first thing to
note is that the high ratio of H-like to He-like line fluxes
is a natural consequence of multi-temperature plasma, since
H-like lines are typically emitted over a wider range of
temperatures than are the He-like lines (see Figure\,\ref{emplot}).

A casual glance at the He-like triplets of Mg and Ne
(Figure\,\ref{triplets}) suggests that the forbidden
component may be present, which would prove a relatively low density
(n$_e < 10^{13}$ cm$^{-3}$), and hence a large (comparable to,
or larger than, the white dwarf) volume.  However, we caution against this
interpretation for several reasons.  First, the observed wavelengths
are not exactly at the laboratory wavelengths of the forbidden transitions,
even though the resonant lines are exactly where they are supposed to be.
This suggest either the identification is incorrect, or that the
forbidden component originates in a material with different
kinematic properties.  Second, observed fluxes in the He-like
triplets of Ne and O are much higher than those of the best-fit
cooling flow model (Figure\,\ref{hetgov}).  This suggests the possibility
of a second component.

The third reason is that the strength of the fluorescent Fe K$\alpha$ line
(equivalent width $\sim$150 eV) is consistent with reflection from
the white dwarf surface and/or the inner disk \citep{DO1997}
that subtends 2$\pi$ steradians, as seen from
the X-ray emitter.  This strongly suggests that the emitting region is
compact.  The likely volume of such a compact boundary layer (much less
than that of the white dwarf) and the emission measure estimates
(Figure\,\ref{emplot}) require that the emission region must be of
very high density.  For a boundary layer covering 1\% of the surface
area of a white dwarf (with radius, $r$, of 6,000 km), and height 0.1$r$,
to have an emission measure of $10^{54}$ cm$^{-3}$, the density must
be $6 \times 10^{14}$ cm$^{-3}$.

In fact, the lack of forbidden components of Mg\,XI and Ne\,IX
at rest wavelengths (see \S 3.3) is consistent with this interpretation.
The ratio $f$/($r$+$i$) is measured to be $<$0.44 for Mg\,XI
and $<$0.16 for Ne\,IX, taking the upper limit for the $f$ flux and
the lower limits for $r$ and $i$.  This implies density greater than
$\sim 4 \times 10^{13}$ cm$^{-3}$ \citep{PD2000}, and can be much higher.

What about the Fe L diagnostics \citep{Maea2001,Maea2003}?  Of the Fe\,XVII
lines, the 17.05\AA\ line appears to be strongly detected, while the
17.10\AA\ line is not confidently detected.  Within the limited S/N
of our data, this is consistent with high density or significant
photoexcitation.  As for the Fe\,XXII lines near 11.8\AA\, the S/N of
our data is not high enough to allow reliable use of this diagnostic,
given the need to first separate the Fe\,XXII 11.78\AA\ line from the
Fe\,XXIII 11.74\AA\ line.

\subsection{Abundances}

Our analysis suggests that the accreting gas in \tgt\ has a high
Ne abundance and a low Fe abundance, relative to the fiducial
values of \cite{AG1989}.  However, note that the solar photospheric
Fe abundance more recently derived by \cite{L2003} is approximately
60\% that of \cite{AG1989}.  Thus, whether the Fe abundance of
\tgt\ is solar or sub-solar may well be a question of the Fe
abundance in the Sun.  However, the Ne overabundance of \tgt\ appears
genuine, regardless of which solar values we adopt.

\tgt\ appears to differ from other CVs for which abundance
anomalies have been reported (see, for example, \citealt{Maea1997,Gea2003}
and references therein), since these concern mostly carbon and nitrogen.
These measurements refer to the abundances of the accreting material,
hence those of the outermost region of the secondary.  The anomalies
are likely caused by an episode of CNO cycle in these binaries,
perhaps in the secondary which was once more massive \citep{Gea2003}.
Although further nuclear evolution can in principle alter the
Ne abundance, this presumably would have been preceded by
the CNO cycle. Since the NV $\lambda$1240 line is seen to be weak
\citep{Fea1982}, the CNO cycle and further nuclear evolution does
not appear to be a viable explanation for the Ne overabundance in \tgt.

An alternative explanation for the CNO abundance anomaly is that
the secondary is contaminated by the ejecta from novae in which CNO cycle
occurred \citep{Gea2003}.  This idea can in principle be extended to
explain the Ne abundance of \tgt, since
approximately a third of all classical novae are thought to harbor an O-Ne-Mg
white dwarf, as evidenced by the high Ne abundances in their ejecta
\citep{Sea1986,Pea1995}.  However, what we observe is not the material
of the underlying white dwarf, nor is it the deeper layers of the accreted
envelop that is enriched by dredge-up.  Instead, the observed X-rays are
emitted above the primary surface by the accreting material.  Thus, for this
mechanism to work, the white dwarf material must first be ejected in
nova eruptions, accreted by the secondary, then transferred back onto
the white dwarf.  This requires a much higher overabundance of Ne in the
nova ejecta than what we observe in the \chandra\ data.
According to the analysis of \cite{P1959}, however, the Ne abundance of the
\tgt\ shell was 1/7,000, or 1.4$\times 10^{-4}$ by number.  This value
is between the Solar value of \cite{AG1989} (1.23$\times 10^{-4}$) and
our X-ray fitting result (1.75$\times 10^{-4}$).  We therefore conclude
that this is not a viable mechanism for the observed Ne overabundance.

This suggests that the \tgt\ system was born with a higher then solar
Ne abundance due to Galactic chemical evolution, implying that it is
either young, or it originated closer to the Galactic center (see, e.g.,
\citealt{Cea1997}).  Given the location of \tgt\ very near the Galactic
mid-plane ($b=0.83^\circ$ at a distance of 360 pc), the former interpretation
is certainly plausible.  It is also consistent with the proposition that
there are two population of classical novae, one young and fast
(\tgt\ being a prominent example), the other older and slower
\citep{DVea1992}.  Abundance measurements of other old novae may provide
a way to test whether there indeed are two distinct populations.

\section{Conclusions}

We have presented our analysis of \chandra, \xte, and archival
X-ray observations of \tgt.  It is among the X-ray brightest
CV and displays a line-rich X-ray spectrum.  However, it is likely
that non-imaging X-ray data of \tgt\ are contaminated by the Galactic
Ridge as well as nearby, unrelated sources.  On \tgt\ itself, our
findings can be summarized as follows.

(1) It is a highly variable X-ray source.  However, we find no indications
of a single, coherent period, and confirm that \tgt\ is a non-magnetic CV.

(2) The X-ray spectrum of \tgt\ is from a multi-temperature, collisionally
ionized plasma.  Specifically, we find that the cooling flow model provides
an adequate framework for physical interpretation.

(3) The spectrum and the luminosity of \tgt\ resembles those of SS~Cyg
in transition, even though the UV luminosity of \tgt\ is very high.
This is a clue that must be taken into account in any detailed physical
models of the X-ray emission mechanism in non-magnetic CVs.

(4) The weak energy dependence of the X-ray flux variations in \tgt\ can
be explained as due to changes in the maximum temperature of
the multi-temperature plasma.  The physical cause of such changes is unknown,
but we note the accretion disk wind in \tgt\ appears to be modulated on
a similar timescale.

(5) There are complications that make the plasma diagnostics less clear-cut
than we would like.  The preponderance of evidence points towards a high
density, compact emission region compatible with the accretion disk boundary
layer.

(6) We detect an apparent overabundance of Ne in the accreting matter.
This is best explained as due to \tgt\ being a young binary.

\acknowledgments

This research has made use of data obtained from the High Energy Astrophysics
Science Archive Research Center (HEASARC), provided by NASA's Goddard Space
Flight Center.



\clearpage

\begin{deluxetable}{lllll}
\tablecaption{Log of Observations}
\tablehead{ \colhead{Observatory} & \colhead{Observation ID} & \colhead{Date}
            & \colhead{Exposure} & \colhead{Average Count Rate} \\
             & & & (ksec) & (\cps) }
\startdata
\asca\    & 34013000 & 1996 October 8--10 & 38(GIS)/ 43(SIS) & 2.22 (total) \\
\xte\     & 30025-01\tablenotemark{a} & 1998 December 5--7 & 41 (total) & 3.00\tablenotemark{b} \\
\xte\     & 60013-01\tablenotemark{c} & 2001 April 17--22 & 55.4 (total) & 3.25\tablenotemark{b} \\
\chandra\ & 1901 & 2001 April 19/20 & 63.5 &  0.33\tablenotemark{d} \\
\enddata
\tablenotetext{a}{Full Observation IDs are 30025-01-01-00, 30025-01-01-01, and
                  30025-01-01-02.}
\tablenotetext{b}{Per PCU. Includes contribution from the Galactic Ridge,
                  estimated to be $\sim$2 \cps\ per PCU.}
\tablenotetext{c}{Full Observation IDs are of the form 60013-01-nn-00,
                  where nn is in the range 01 to 20.}
\tablenotetext{d}{$\pm$1st orders combined.}
\label{obstab}
\end{deluxetable}

\begin{deluxetable}{cccccl}

\tablecaption{Global Fit Results}
\tablehead{ \colhead{N$_{\rm H}$} & \colhead{kT$_{\rm min}$} &
            \colhead{kT$_{\rm max}$} & \colhead{Ne\tablenotemark{a}} &
            \colhead{Fe\tablenotemark{a}} &
            \colhead{$\dot M$\tablenotemark{b}} }
\startdata
$< 1.3 \times 10^{19}$ cm$^{-2}$ & $<$ 0.14 keV &
20.3$^{+1.1}_{-1.8}$ keV & 1.42$^{+0.21}_{-0.18}$ & 0.62$^{+0.07}_{-0.05}$ &
8.00$\pm 0.01 \times 10^{-11}$ M$_\odot$\,yr$^{-1}$ \\
\enddata
\tablenotetext{a}{Abundances, relative to solar values of \citet{AG1989}}
\tablenotetext{b}{Participating mass accretion rate, for a fiducial
                  distance of 360 pc \citep{HD1927}.}
\label{gfres}
\end{deluxetable}

\begin{deluxetable}{lllll}
\tablecaption{Emission Lines}
\tablehead{ \colhead{Line ID} & \colhead{Wavelength}
	& \colhead{Width} & \colhead{Flux} \\
	 & \colhead{(\AA)} & \colhead{(km\,s$^{-1}$)} &
	\colhead{(10$^{-6}$ photons\,cm$^{-2}$s$^{-1}$)} }
\startdata
Fe\,XXVI 1.78 & 1.774 (1.767--1.787) & 1150 ($<$3050) & 12.5 (4.9--24.2) \\
Fe\,XXV 1.86 & 1.855 (1.851--1.860) & 550 ($<$2250) & 24.3 (15.2-40.3) \\
Fe Fluorescent & 1.939 (1.933--1.945) & 1520 (770--2680) & 20.9 (12.5-31.3) \\
S\,XVI 4.73 & 4.732 (4.713--4.738) & 160 ($<$660) & 5.9 (2.6-10.1) \\
S\,XV 5.04 & 5.043 (5.034--5.048) & 0 ($<$800) & 4.9 (2.8--10.5) \\
Si\,XIV 6.18 & 6.182 (6.180--6.184) & 300 (180--430) & 19.3 (16.4--22.7 ) \\
Si\,XIII 6.65 & 6.651 (6.647--6.659) & 460 (230--970) & 12.0 (8.7--15.7) \\
Fe\,XXIV 7.99 & 7.981 (7.976--7.990) & 370 (10--750) & 3.8 (2.0--6.0) \\
Mg\,XII 8.42 & 8.420 (8.418--8.422) & 280 (200--400) & 18.0 (15.3-21.1) \\
Mg\,XI 9.17 & 9.170 (9.167--9.175) & 0 ($<$180) & 5.7 (4.0--8.4) \\
Mg\,XI 9.23 & 9.230 (9.220--9.237) & 0 ($<$180) & 2.8 (1.2--5.2) \\
Mg\,XI 9.31 ?\tablenotemark{a} & 9.335 (9.326--9.340) & 0 ($<$600) & 3.2 (1.1--5.9) \\
Fe\,XX 10.00 & 9.978 (9.975--9.988) & 0 ($<$370) & 6.6 (4.1--9.6) \\
Fe\,XX 10.05 & 10.051 (10.041--10.061) & 0 ($<$360) & 2.8 (1.1--5.7) \\
Ne\,X 10.24 & 10.245 (10.240--10.254) & 0 ($<$340) & 6..4 (3.6--9.7) \\
Blend\tablenotemark{b} & 10.644 (10.637--10.653) & 660 (480--990) & 24.0 (18.0--30.8) \\
Blend\tablenotemark{c} & 11.019 (11.010--11.029) & 590 (350--820) & 18.7 (13.4--25.1) \\
Fe\,XXIV 11.18 & 11.177 (11.170--11.183) & 400 (230--690) & 16.1 (11.4--21.9) \\
Fe\,XXIV 11.43 & 11.439 (11.429--11.440) & 0 ($<$280) & 10.1 (6.8--15.3) \\
Blend\tablenotemark{d} & 11.756 (11.737--11.773) & 820 (510--1280) & 15.4 (8.3--23.8) \\
Ne\,X 12.13 & 12.135 (12.132--12.139) & 270 (120--360) & 45.6 (36.2--53.4) \\
Ne\,IX 13.45 & 13.439 (13.431--13.445) & 220 (100--380) & 19.8 (14.2--30.7) \\
Ne\,IX 13.55\tablenotemark{e} & 13.531 (13.516--13.544) & 800 (580--1130) & 37.3 (27.2--48.3) \\
Ne\,IX 13.70 ?\tablenotemark{f} & 13.744 (13.730--13.756) & 70 ($<$570) & 7.5 (2.5--12.3) \\
Fe\,XVII 14.37 & 14.360 (14.355--14.368) & 0 ($<$280) & 8.8 (4.7--16.1) \\
Fe\,XVII 15.01 & 15.009 (15.001--15.014) & 270 (10--480) & 46.3 (23.7--48.9) \\
Fe\,XVII 15.26 & 15.260 (15.256--15.267) & 0 ($<$190) & 15.5 (11.3--26.5) \\
Fe\,XVIII 15.63 & 15.623 (15.607--15.641) & 300 (100--630) & 8.0 (4.4--20.7) \\
Fe\,XVIII 16.05 & 16.035 (15.996--16.054) & 760 (460--1040) & 31.3 (19.2--45.4) \\
Fe\,XVII 16.78 & 16.780 (16.764--16.789) & 370 (260--650) & 30.6 (19.2--44.0) \\
Fe\,XVII 17.05 & 17.061 (17.052--17.071) & 440 (340--650) & 49.9 (35.5--67.1) \\
O\,VIII 18.97 & 18.964 (18.958--18.970) & 330 (270--430) & 144.2 (115.6--169.3) \\
\enddata
\label{llist}
\tablenotetext{a}{Flux at the rest wavelength of the forbidden Mg~xi is
                  $< 2.8 \times 10^{-6}$ photons cm$^{-2}$s$^{-1}$}
\tablenotetext{b}{Blend of Fe\,XIX 10.64 and Fe\,XIX 10.65?}
\tablenotetext{c}{Blend of Fe\,XXIII 11.02 and Fe\,XXIV 11.03?}
\tablenotetext{d}{Blend of Fe\,XXIII 11.74 and Fe\,XXII 11.77?}
\tablenotetext{e}{Possibly blended with Fe,XIX 13.52}
\tablenotetext{f}{Flux at the rest wavelength of the forbidden Ne~ix is
                  $< 3.0 \times 10^{-6}$ photons cm$^{-2}$s$^{-1}$}
\end{deluxetable}

\end{document}